\documentstyle[aps,prl,float,epsfig]{revtex} 

\begin{document}
\draft
\preprint{HEP/123-qed}

\wideabs{
\title{\Large\bf Evidence for topological nonequilibrium in magnetic
configurations}
\author{\normalsize{S. I. Vainshtein$^1$, Z. Miki\'c$^2$, R.
Rosner$^1$, and J. A. Linker$^2$}\\
{\small\it $^1$Department of Astronomy and Astrophysics, University
of Chicago, Chicago, Illinois 60637}\\
{\small\it $^2$Science Applications International Corporation, San Diego,
California 92121}}
\date{\today}
\maketitle
\begin{abstract}
\normalsize{
We use direct numerical simulations to study the evolution, or relaxation, of
magnetic configurations to an equilibrium state. We use the full
single-fluid equations of motion for a magnetized,
non-resistive, but viscous fluid; and a Lagrangian approach is used to
obtain exact solutions for the magnetic field. As a result, the topology of
the magnetic field remains unchanged, which makes it possible to study the case
of topological nonequilibrium. We find two cases for which such nonequilibrium
appears, indicating that these configurations may develop singular current
sheets. }
\end{abstract}
\pacs{PACS number(s): 52.30.Bt, 52.30.-q, 47.65.+a}
}

\narrowtext

\section{Introduction.}
\label{I}

Formation of singularities, or current sheets, is one of the striking
features of
astrophysical as well as tokamak plasmas \cite{plasma}. Such singularities
are key
to understanding active phenomena related to fast magnetic field reconnection
\cite{flares}, \cite{golub}. For example, fast dynamos rely on fast
reconnection
of magnetic field lines \cite{priest}, \cite{dynamo}. Despite their importance,
key issues related to current sheet formation are still not well understood.
Supposing, e.g., that they are formed due to instabilities, one has to
assume that
fluid dynamical processes are able to slowly deform equilibrium magnetic field
configurations (and thereby build up regions of field gradients) without
significant reconnection until a marginal state is reached. At this threshold,
instability-driven reconnection would then lead to release of the stored free
energy on the (observed) time scales thought to be too short to be consistent
with, for example, Sweet-Parker reconnection \cite{golub}, \cite{priest}.
However,
it has been long recognized \cite{galeev} that in the presence of
reconnection, it
is not obvious how one can attain (meta)stable configurations which store
significant free energy. Furthermore, it is not clear why reconnection
would not
simply return the system to the marginal state, thus releasing only a small
fraction of the available free energy.

In this paper, we explore one possible solution to these puzzles: We consider
specific magnetic field configurations which could arise from a slow evolution
of (stable) quasi-equilibria, and then examine their subsequent (unforced)
evolution. Our aim is to show that there exist configurations that evolve
initially on the slow rate, but that can reach a point at which spontaneous
current sheet formation occurs. These configurations have been referred to as
``topological nonequilibria" (TN) \cite{flares}, \cite{parker}, and lead to
situations in which the topology of the field is such that in a relaxed
equilibrium state it inevitably contains discontinuities. TN results in
spontaneous reconnection, because no external forces are involved; and in the
cases we shall examine, the result is that extraction of all of the
available free
energy becomes possible.

Finally, we note that an important aspect of this problem relates to the fact
that there is a direct correspondence between magnetostatic equilibria and
steady Euler flows, as pointed out by Moffatt \cite{analogy}; this
problem is therefore closely connected to the possible formation of
singularities in hydrodynamics; see also \cite{more}, \cite{Bajer}.

\section{Description of the approach.}
\label{II}

\subsection{The idea of topological nonequilibrium.}
\label{2a}

The main ideas of topological nonequilibrium (henceforth, TN) were formulated
rigorously by Moffatt \cite{analogy}, \cite{more}, \cite{Bajer}. Consider an
ideally conductive viscous (ICV) flow. We restrict ourselves to incompressible
flows. Starting with initial magnetic field ${\bf B}({\bf x}, t=0)={\bf
B}_0({\bf x})$ of arbitrary topology, one expects that such a configuration
will
relax to a static state, with zero velocity field, and nontrivial magnetic
field
${\bf B}_E$. The latter configuration is then called `topologically accessible'
because the field's topology does not change during this frozen-in evolution.
If this relaxed equilibrium state contains discontinuities, then all of the
states in the evolution are referred to as TN. It may be expected in a
realistic situation, when small but finite resistivity $\eta$ is taken into
account, that these discontinuities evolve into finite--width current sheets,
resulting in efficient reconnection and dissipation of the magnetic field.
Unfortunately, there are only a few special cases for which it is possible to
demonstrate that TN exists \cite{kadom}. In this paper, we restrict
ourselves to analysis of two-dimensional configurations, and study the
evolution
of two generic field configurations which can lead to TN.

Of course, in general, there is no reason that a given initial configuration is
at equilibrium. However, one would normally expect that, after relaxation,
such a configuration would evolve to attain a smooth equilibrium, and that the
magnetic field evolution subsequently stops. Some initial field topologies,
however, cannot possibly relax to a smooth equilibrium, resulting in TN. It is
obvious that use of the word ``nonequilibrium" is not strictly correct,
because in
the final state the field {\it is} at equilibrium as long as the diffusivity
vanishes exactly. However, in the spirit of maintaining already existing
tradition, we retain this terminology.

As a result of  relaxation, the magnetic field will reach an equilibrium
state. In two dimensions, $B_x=\partial_y A$, $B_y=-\partial_x A$, and the
flux $A$ obeys in equilibrium the equation
\begin{equation}
\nabla^2 A=-4\pi \frac{dP(A)}{dA},
\label{3}
\end{equation}
where $P=p+B_z^2/(8\pi)$, $p=p(A)$, and $B_z=B_z(A)$; see, e.g., \cite{book}.
As an aside, we note that if the pressure $p$ can be neglected, then this
equilibrium is force-free; this may occur in specific applications such as in
the solar corona. In addition,  the total pressure $P$ (Eq.\ (\ref{3}))
ought to substantially exceed the transverse magnetic energy
$B_x^2+B_y^2$ in order to justify the incompressibility assumption for
the evolution to this equilibrium state.

Equation (\ref{3}) is trivially satisfied in the one-dimensional case. To
start with, suppose that $A$ is a function of $x$ only, corresponding to
straight field lines parallel to $y$ axis. An initial arbitrary distribution
is generally not at equilibrium. However, after relaxation, the field
reaches the well-known equilibrium
\begin{equation}
\frac{B_y(x)^2}{8\pi}+P(x)=\rm const,
\label{4}
\end{equation}
automatically satisfying (\ref{3}). The same is true for axisymmetric
configurations, when $A=A(r)$, and the field consists of concentric
circles.
$$
\psfig{file=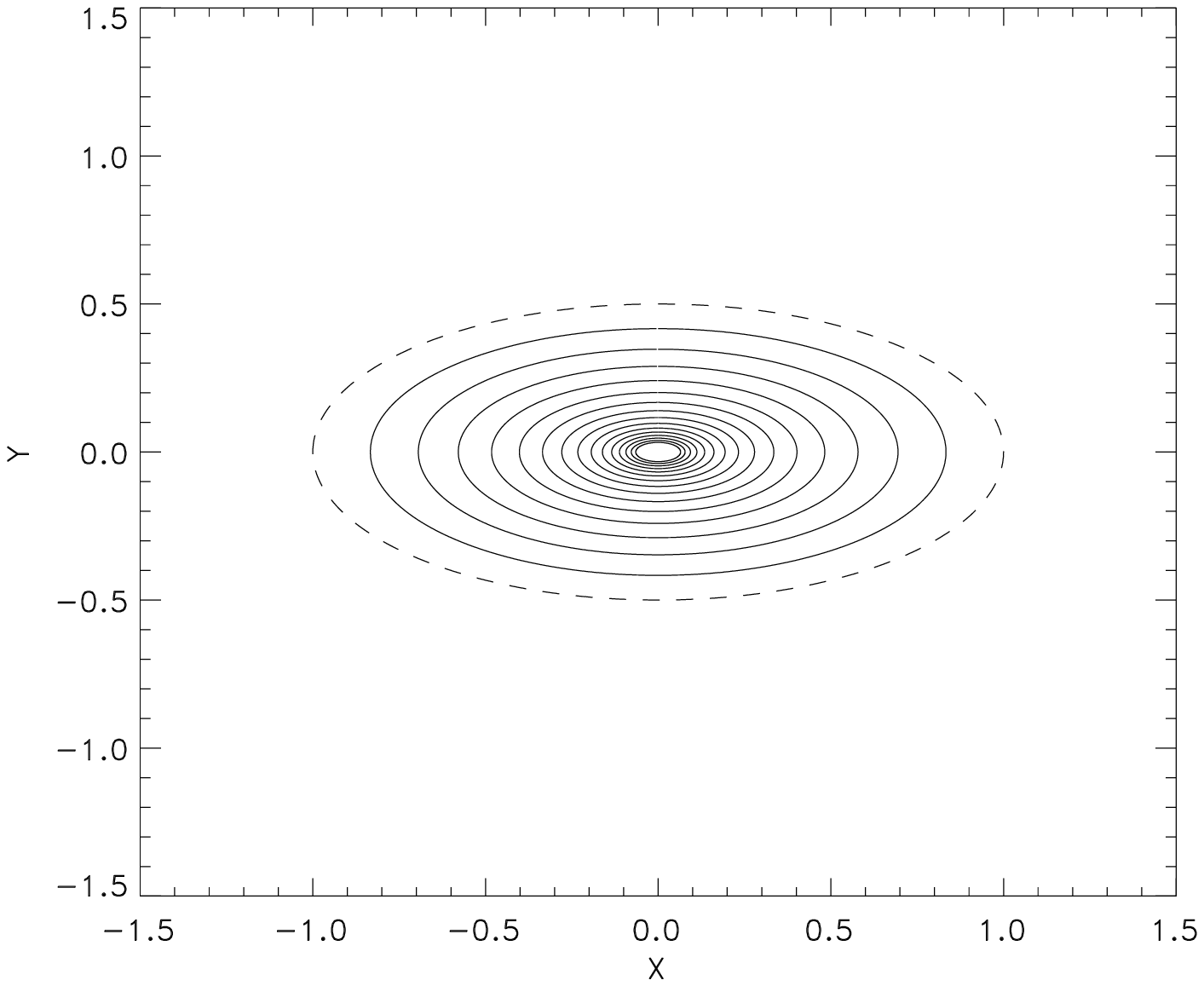,width=3.5in}
$$
\begin{figure}
\caption{Sketch of the type A topology configuration: closed nested
magnetic field
lines. The dashed line here, as in all the figures below corresponds to a
field line with vanishing magnetic field strength.}
\end{figure}
Going to two dimensions complicates the problem considerably. Consider
first a configuration with closed nested field lines, so that $A(x,y)$
has one maximum (minimum). The configuration is depicted in Fig.\ 1, and
we will refer to it below as ``case A". Of course, the axisymmetric
configuration is topologically accessible from this configuration, and
therefore it can reach equilibrium. The question, however, is if this
equilibrium is unique.

$$
\psfig{file=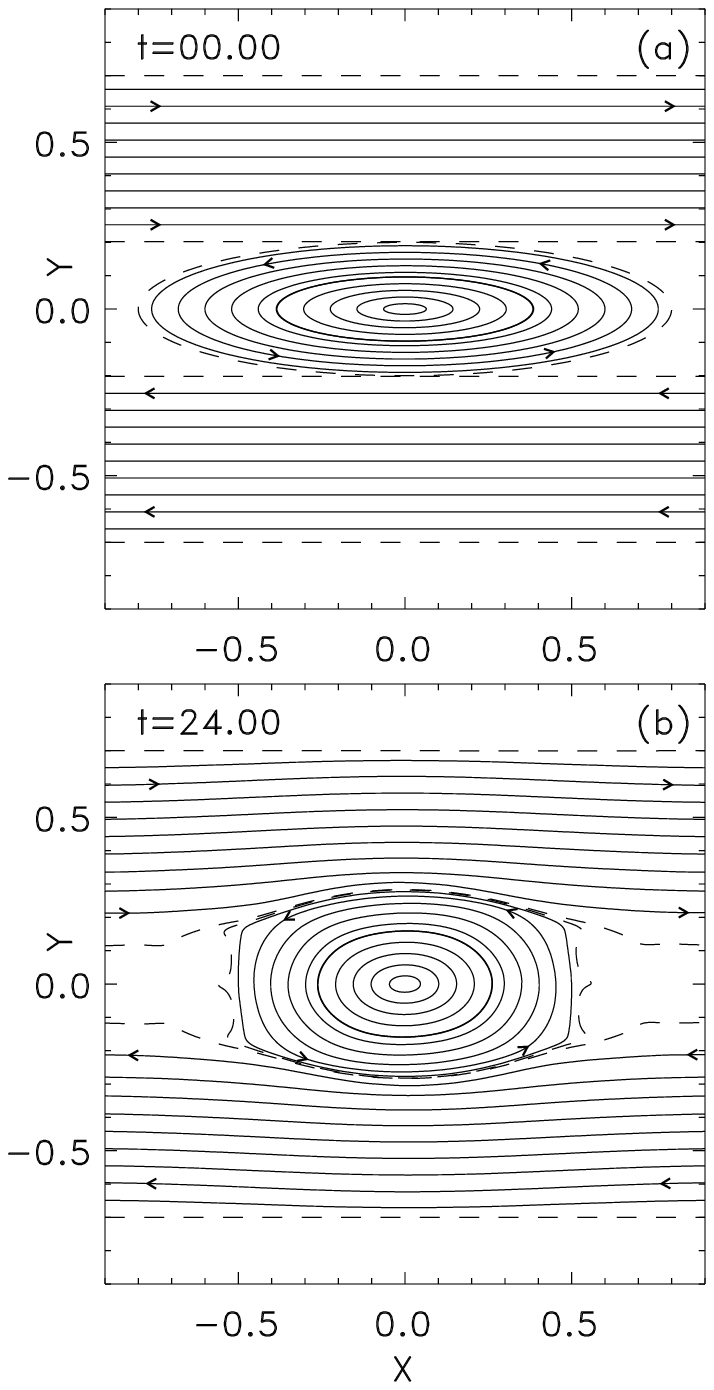,width=3.5in}
$$
\begin{figure}
\caption{Ellipse-shaped configuration placed between two horizontal magnetic
walls; the initial configuration was generated by considering two families of
(parametric) curves, i.e., straight lines and ellipses, and joining them as
shown
in panel (a). As a result of evolution of the configuration shown in panel
(a), the
field evolves such that it is pushed to the walls to form discontinuities, as
shown in panel (b). Dashed lines correspond to ${\bf B}_\perp=0$. 
Both panels present
results of numerical simulations, but only selected field lines are
depicted for
illustrative purposes.}
\end{figure}
If there exists a magnetostatic equilibrium with type A field topology
with essentially arbitrary field line geometry, e.g., with elliptic field
lines, as in Fig.\ 1, then we would expect an arbitrary type A configuration to
relax to this equilibrium without dramatic changes in its geometry. However,
suppose this equilibrium exists only in axisymmetric form, i.e., can only be
realized with concentric (field line) circles; then, if this configuration were
placed between magnetic ``walls" (such as regions of strong magnetic fields
in the solar corona, or solar wind), as in Fig.\ 2a, we would expect the
formation
of discontinuities (because it  would not be possible to evolve to the
equilibrium
state).  Furthermore, if we allowed for nonvanishing diffusion, then such a
configuration would not settle down until {\it all} magnetic lines are
reconnected, and the bubble seen in Fig.\ 2 disappears entirely.

It is useful to expand slightly on the astrophysical relevance of this case.
Our point is that ``case A" shown in Fig. 2a can be regarded as an
abstraction of a
commonly expected field configuration in the solar atmosphere: Consider the
emergence of a magnetic flux tube from the solar interior to the corona,
where it
enters a highly conducting medium already suffused by pre-existing magnetic
fields. If one abstracts such an emerging flux tube as a rising cylinder,
then the
expected field topology in planes perpendicular to the tube axis should be
similar
to case A: The nested closed field lines in such planes then represent the
toroidal field component of the emerging flux tube; and the magnetic ``walls"
shown in Fig. 2a represent the projections in such planes of the magnetic
fields
of the surrounding magnetized coronal plasma.

$$
\psfig{file=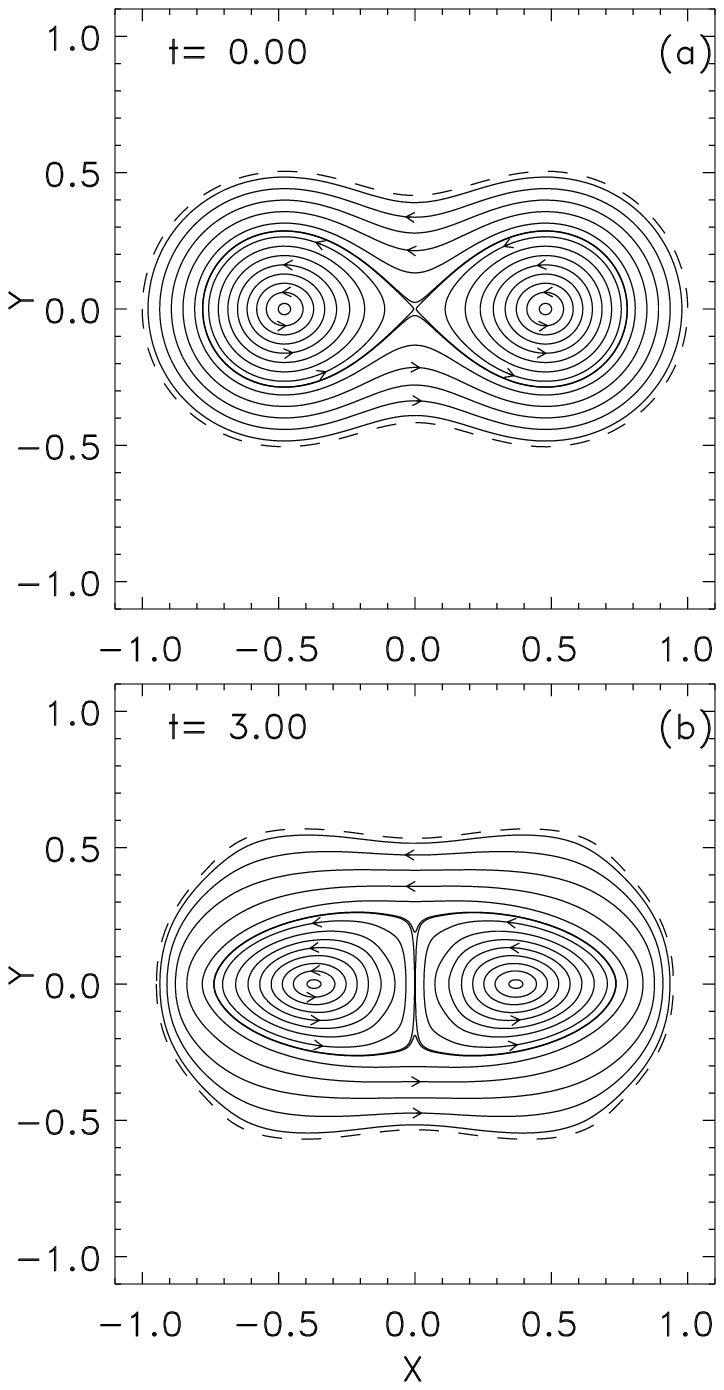,width=3.5in}
$$
\begin{figure}
\caption{Simulations for the (initially continuous) rosette structure, as
depicted in panel (a). As a result of the relaxation, this configuration
evolves into a field containing a discontinuity between the two magnetic
islands,
as shown in panel (b). The presence of the external zero line (dashed line) is
vital  for TN.}
\end{figure}
The second type of field topology we consider below is what we call type B;
this
more complicated topology is a ``rosette structure" (Fig.\ 3a), which has been
investigated experimentally \cite{Yamada}. In terms of the flux function
$A(x,y)$, this configuration consists of two maxima, e.g., two ``mountains",
surrounded by a pedestal, i.e., two magnetic islands surrounded by closed
magnetic field lines going around the two islands; the field vanishes outside
the zero-line. If the type B topology cannot exist in smooth equilibrium,
then a
current sheet develops, resulting in efficient reconnection of field lines
until all field lines of the islands are reconnected, and eventually only one
island remains, of the topology of the type A. In contrast, if this kind of
topology does exist in smooth equilibrium, then nothing dramatic would happen,
and the configuration would relax to this equilibrium without any
discontinuities. The astrophysical context in which this type of configuration
may be created is similar to that just described above: consider the
emergence of
two adjacent twisted solar flux tubes into a non-magnetized ambient corona;
again, the field structure in a cross-section perpendicular to the tube
axes will
appear as shown in Fig. 3a. Thus, in both cases A and B, we are dealing
with the
generic case of bounded magnetic flux systems (i.e., systems of magnetic field
lines which lie within a finite bounding surface on which the field vanishes),
which can be regarded as abstractions of, for example, isolated flux tubes
emerging into the highly conductive solar corona.

Generally, finding TN states is far from trivial. To illustrate, let us
return to the type A topology. The axisymmetric equilibrium solution is
{\it not}
unique. For example, one can construct a solution to (\ref{3}),
$$
A(x,y)=\sin{kx}\sin{ky},
$$
depicted in Fig.\ 4a, which has the same topology of field lines as depicted in
Fig.\ 1. This asymmetric field is at equilibrium, so that the general answer to
the question of whether, say, elliptic configurations of the form shown in
Fig.\ 1 can be at equilibrium, is affirmative. An analogous construction can be
carried our for the topology of type B; in this case, the solution of (\ref{3})
can be constructed as
$$
A_z=\sum_n A_n e^{ik_x^{(n)}x+ik_y^{(n)}y},
$$
where $(k_x^{(n)})^2+(k_y^{(n)})^2= \rm constant$ (see, e.g., \cite{i},
\cite{parker}). This example is depicted in Fig.\ 4b; the rosette structure
shown is at equilibrium without any discontinuities.

The situation changes if a zero-line (a line where ${\bf B}_\perp=0$,
${\bf B}_\perp=\{B_x,B_y\}$,  and generally $B_z\not= 0$) is present,
such as the dashed line shown in Fig.\ 1 for the type A, and in Fig.\ 3 for the
type B field topology. The zero-line possess two remarkable properties.
First, the
magnetic field remains zero on this line in the presence of ICV flow. Thus, if
we write the ideal induction equation in the form,
$$
\frac{ d{\bf B}}{dt}=({\bf B}\cdot\nabla){\bf v}~,
$$
which in 2D reads,
\begin{equation}
\frac{ d{\bf B}_\perp}{dt}=({\bf B}_\perp\cdot\nabla){\bf v}, ~~~
\frac{ dB_z}{dt}=0,
\label{1}
\end{equation}
it is easy to see that because the left-hand side describes transport of
any fluid element (in particular, of the zero-line) by the motion, and because
the right-hand side corresponds to change of the field along the Lagrangian
trajectory (and as the right-hand side vanishes on the zero-line), this
equation will preserve the property ${\bf B}_\perp=0$ on the zero-line.

$$
\psfig{file=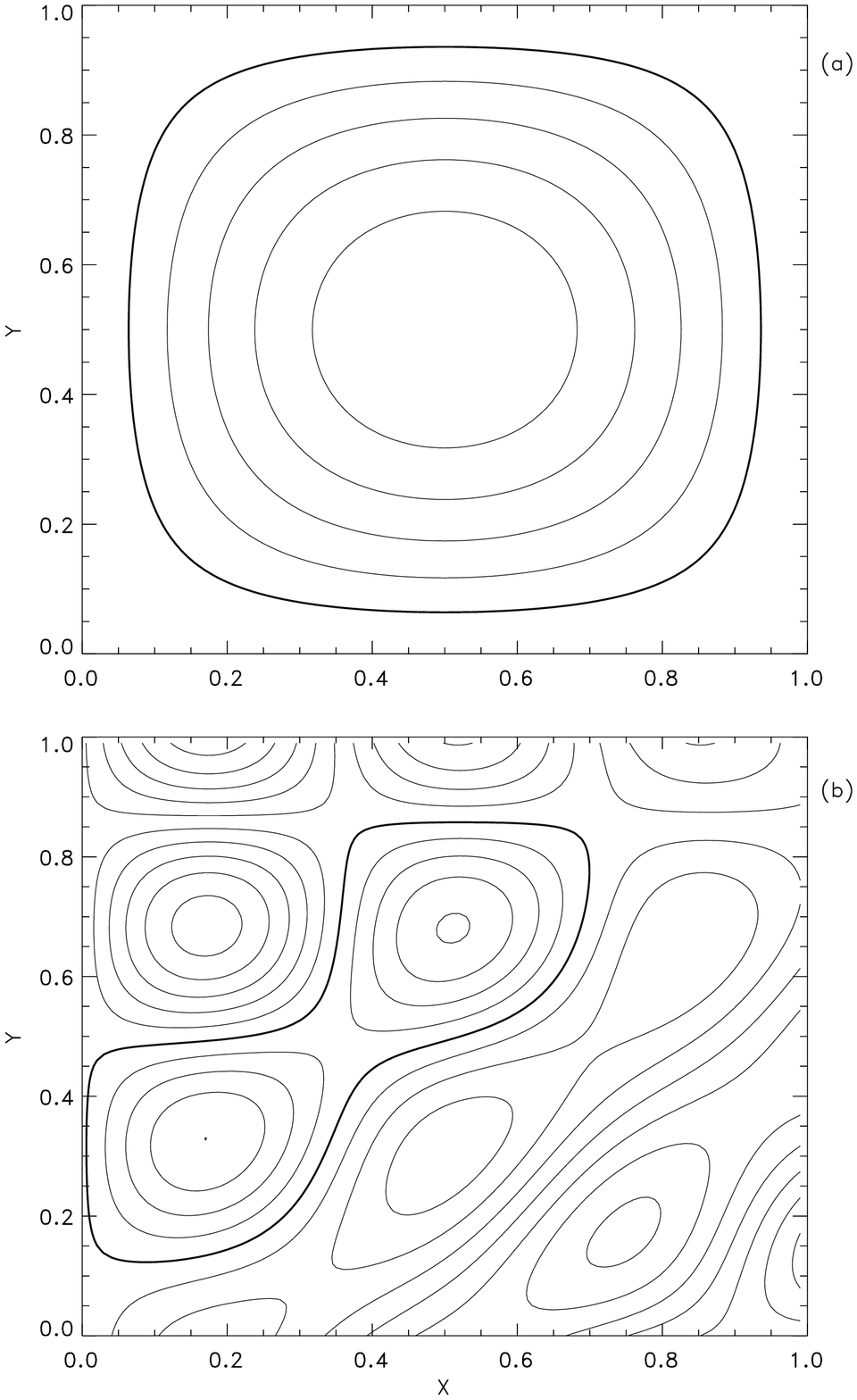,width=3.5in}
$$
\begin{figure}
\caption{Two examples of field configurations which share the field topology of
the two cases (A and B) we are studying, which do not have any simple
symmetries, but are nevertheless smooth equilibria. (a) Type A (marked by the
thick line); (b) Type B (rosette structure, marked by a thick line).}
\end{figure}
Second, if the zero-line has a constant (along the line) curvature, e.g., is a
straight line or a circle, and if the field is also analytical, then the entire
configuration will have the same geometry as the zero-line. In other words,
if the zero-line is a straight line, then all other field lines are straight as
well; alternatively, if the zero-line is a circle, then the analytical
equilibrium configuration consists of concentric circles. The proof is easily
constructed by expanding $A(x,y)$ in the vicinity of the zero-line \cite{book}.
Note, however, that the constant curvature zero-line is a special case
(although
it can be regarded as a representation of the emergence of magnetic flux
on, for
example, the solar surface, in which geometrically symmetric flux bundles
straddle the separating ``neutral line"); in general, the zero-line is
arbitrary
in shape, as shown in Figs. 1 -- 3. Nevertheless, we may conjecture that the
zero-line imposes a severe constraint  on the geometry: That is, we conjecture
that the existence of this line results in unique (smooth) solutions of the
equation (\ref{3}) in the form of magnetic field lines with constant curvature
\cite{book}.

One of the considerations in favor of this conjecture is as follows.
Without loss of generality, $A\equiv 0$ outside the configuration, and thus
$A=0$
on the boundary (whose shape is as yet unspecified), corresponding to the
Dirichlet
problem for equation (\ref{3}). On the other hand,  because $B_x=B_y=0$ on the
same boundary, we have $\partial_n A=0$, corresponding to a Neumann
problem. The problem is thus over-constrained; and one would expect this to
lead to
degeneracy of the solution. That is, these specific boundary conditions are
expected to restrict the shape of the boundary itself, and thus in turn to
restrict the topology of possible equilibria. Although the boundary
conditions are specified, and the problem is thus rigorously formulated,
the above
statement regarding the over-constrained nature of our problem nevertheless has
not been shown to be useful in constructing a formal mathematical proof
concerning
the geometry of the configuration in the presence of a zero-line. Formally,
it is
easier to discard the TN for a given topology by direct construction of a
solution with needed properties. Generally, it is not clear at all how to
construct a formal proof that the only solution of (\ref{3}) with boundary
condition ${\bf B}_\perp=0$ for the type A topology is unique, and axisymmetric, thus
defining the shape of the boundary itself! It is even less clear how to
prove that there is no smooth solution for the type B topology, assuming
that this
statement is true.

Finally, we note that one can simply produce artificial discontinuities, but
these are irrelevant to our discussion. To illustrate, suppose that we place
superconductive walls at the locations of the thick lines in both panels of
Fig.\ 4. In that case, the configurations will be at equilibrium (for both
type A
and B topologies), and the field will be smooth everywhere except at the
boundaries (where the field jumps from a finite value just outside the walls to
zero at the walls in order to meet the boundary condition of zero field
within the
superconducting walls). This type of discontinuity is irrelevant to the
astrophysical problem we are aiming at, and we therefore do not discuss it any
further.

\subsection{Description of the solution method.}
\label{2b}

One of the powerful ways to study the formation of current sheets is via
numerical simulation. However, in numerical simulations the Lundquist number,
$S=c_A L/\eta$ ($c_A$ the Alfv\'en speed, and $L$ the characteristic length),
which is critical for this problem, is far below that corresponding to
values encountered in natural systems, viz., under astrophysical conditions
\cite{golub}. When $S$ is not sufficiently large, the separation between
typical reconnection times and typical fluid dynamical times may not be 
large; it
is therefore difficult   interpret realistic resistive calculations in
the
context of a problem in which current sheet formation is to occur without
topological changes.
On the other hand, numerical schemes which attempt to
circumvent this problem by solving the ideal MHD equations suffer from the
difficulty that such schemes may be subject to numerical
instability, so that it becomes difficult to distinguish between numerical
artifact and physically correct current sheet formation. When
discontinuities in
the magnetic field appear, traditional numerical MHD codes tend to either break
down, or to introduce a small amount of resistivity to broaden the current
sheets
(so that, for example, their width is larger than a mesh cell). In certain
situations, the symmetry of the problem can be exploited to study the
approach to
the ideal solution, i.e., $\eta\to 0$, see, e.g., \cite{zmjl}.  However,
typical
simulations actually add some amount of numerical resistivity, as in
\cite{Bajer},
so that numerical solutions of the ideal induction equation correspond to
solutions of that equation with an added effective diffusivity. 
In studies of reconnection it is known that the specification
of boundary conditions on the magnetic field and velocity
(which specify the rate at which magnetic field and plasma
is brought into the reconnection layer) may affect
the rate of reconnection.
In our simulations we study spontaneous formation of singularities
by isolating the flux system from the boundaries.
We surround our flux bundle by a (transverse)
field-free region, and we place the boundaries far away
from the bundle, thereby minimizing the effect
of the boundary conditions on the formation of the current sheets.

We address this issue as a relaxation problem in the framework of ICV
flows. Our approach involves a direct numerical simulation of ICV flows, i.e.,
solving the set of equations (\ref{1}) and the momentum equation,
$$
\frac{d{\bf v}}{dt}=\frac{\partial {\bf v}}{\partial t}
+({\bf v}\cdot\nabla){\bf v}=
$$
\begin{equation}
=-\frac{1}{\rho}\nabla p + \frac{1}{4\pi\rho}\{\nabla
\times {\bf B}\}\times {\bf B}+\nu\nabla^2 {\bf v},
\label{2}
\end{equation}
with $\nabla\cdot {\bf v}=0$ \cite{porous}. We use a Lagrangian approach to
solve the induction equation (\ref{1}), as in \cite{suppression},
\cite{dynamo}.
More specifically, the magnetic field inside the region of interest is
represented by a large number of field lines; the evolution of the field
lines is
then followed using the exact Lundquist solution, i.e., knowing the initial
strength of magnetic field on a fluid element connecting two nearby points on a
field line, the final strength is proportional to the length of the
segment, as it
is stretched by the motions. We assume for all cases that the magnetic field
vanishes on the outermost field line (the dashed curves shown in the figures).
The number of field lines which fill the domain is chosen so that the
subsequent field evolution can be followed without leaving gaps in the
final state,
i.e., we determine the number of initial field lines by fixing the spatial
resolution of the final state; we discuss this point further immediately
below. As
an important aside, we note that the initial magnetic field is smooth, implying
that the current system, ${\bf j} (x,y)$, which is defined by Ampere's law
$$
\nabla \times {\bf B} =\frac{4\pi}{c}{\bf j},
$$
is smooth as well, i.e., there are no current sheets initially.

The momentum equation (\ref{2}), in contrast, is solved using standard finite
difference techniques, with finite viscosity. However, the requirement of
coupling the magnetic field evolution to the momentum equation does lead to a
complication for computing the Lorentz force. The key issue is that the
momentum equation requires the Lorentz force to be evaluated on a homogeneous
spatial grid, while the magnetic field evolution is given in Lagrangian
space. We
resolve this issue by (quadratically) interpolating the Lorentz force at
each time
step onto the homogeneous grid used by the momentum equation (\ref{2});
similarly,
we use quadratic interpolation from the momentum equation mesh to evaluate the
velocity field on the Lagrangian mesh. In order to minimize interpolation
errors,
we fix the number of field lines such that every Eulerian grid domain is
pierced
by at least a few field lines throughout the calculation. Note here that
interpolation errors do lead to inaccuracies in the solution of the flow and
magnetic fields, but by construction cannot lead to changes in the magnetic
field
topology. Note also that we have checked for convergence of the solutions
as the
spatial resolution of our calculation is increased; our conclusion is that the
results presented here do not depend on grid resolution. Our solution
corresponds
to the limit $\eta\to 0$, in the sense that the topology is strictly conserved,
but with finite viscosity; thus, the computational scheme we use forces the
relaxation to be due solely to viscous damping, and as a consequence, the field
relaxes to an equilibrium state. Our approach has the dual virtues that the
boundary conditions for the magnetic field do not need to be specified, 
and that the field
topology is preserved; it is therefore appropriate for the study of TN.

If the viscosity is large, then (\ref{1}-\ref{2}) describe monotonic
relaxation to equilibrium. We can estimate the relaxation time as follows: from
(\ref{2}) we find that $v \sim c_A^2 L/\nu=c_A S_\nu$, where $S_\nu=c_AL/\nu$.
This viscous regime is realized if $S_\nu \ll 1$. The relaxation time is then
$t_\nu \sim L/v=\tau_A/S_\nu$, with $\tau_A=L/c_A$. In the opposite
limiting case,
$S_\nu \gg 1$, the system undergoes (strong) Alfv\'en oscillations ($v\approx
c_A$), with a period $\tau_A$, decaying on a viscous time $t_\nu \sim
L^2/\nu=\tau_A S_\nu$. These two cases can be jointly described by an
interpolation formula,
$$
t_\nu=\tau_A (S_\nu +1/S_\nu),
$$
from which it follows that the relaxation time is large for both limiting
cases (in terms of $\tau_A$). Thus, optimal relaxation to equilibrium
occurs for $S_\nu \sim O(1)$; in the simulations, we used the value $S_\nu=5$.
It is important that $S_\nu$ not be too large: An important constraint on the
value of $S_\nu$ is that the simulations remain stable. This constraint is not
met if $S_\nu$ is too large; because the two dynamical equations are  solved in
different coordinates [eq.\ (\ref{1}) in Lagrangian, and eq.\ (\ref{2}) in
Eulerian coordinates], errors arise from the interpolation from one coordinate
system to the other, and therefore the calculations make sense only if these
errors are damped sufficiently by viscosity.

\section{Description of the results.}

We conducted two series of numerical experiments for case A.  In the first
series, we consider the relaxation of this type of topology without any
external
field, as depicted in Fig.\ 1. We explored different initial shapes of the
field
lines, including ellipse-like, diamond-like, and other similar
configurations. In
addition, for a fixed shape, we explored different distributions of the flux
function $A(x,y)$, i.e., different functional dependences $A(s)$, where $s$
labels the field lines. The results are always the same: the field ends up
in an axially symmetric
state, provided the field vanishes on the outermost field line.

In another sets of experiments, this same configuration (case A) is placed
between magnetic walls, as in Fig.\ 2a; in this case, the system always evolves
to create discontinuities, as in Fig.\ 2b, where the field lines are taken from
one of our simulation runs. We see that as the ``bubble" evolves, it
attempts to
become axisymmetric, but as it does so, two discontinuities begin to form, as
depicted in Fig.\ 2b, see also Fig. 5a. It is interesting to note that, 
for some initial
conditions, a {\it current  point} is formed, rather than a current sheet (or a
line in two dimensions), suggesting that finite conductivity could presumably
result in fast reconnection; that is, according to the Sweet-Parker mechanism
(see, e.g., \cite{priest}), the reconnection rate $v_d \sim 1/\ell$, where
$\ell$
is the length of the current sheet, so that a short current sheet speeds up the
reconnection. (In the classical Sweet-Parker mechanism, $\ell=L$, and
$v_d=c_A/S^{1/2}$.)

$$
\psfig{file=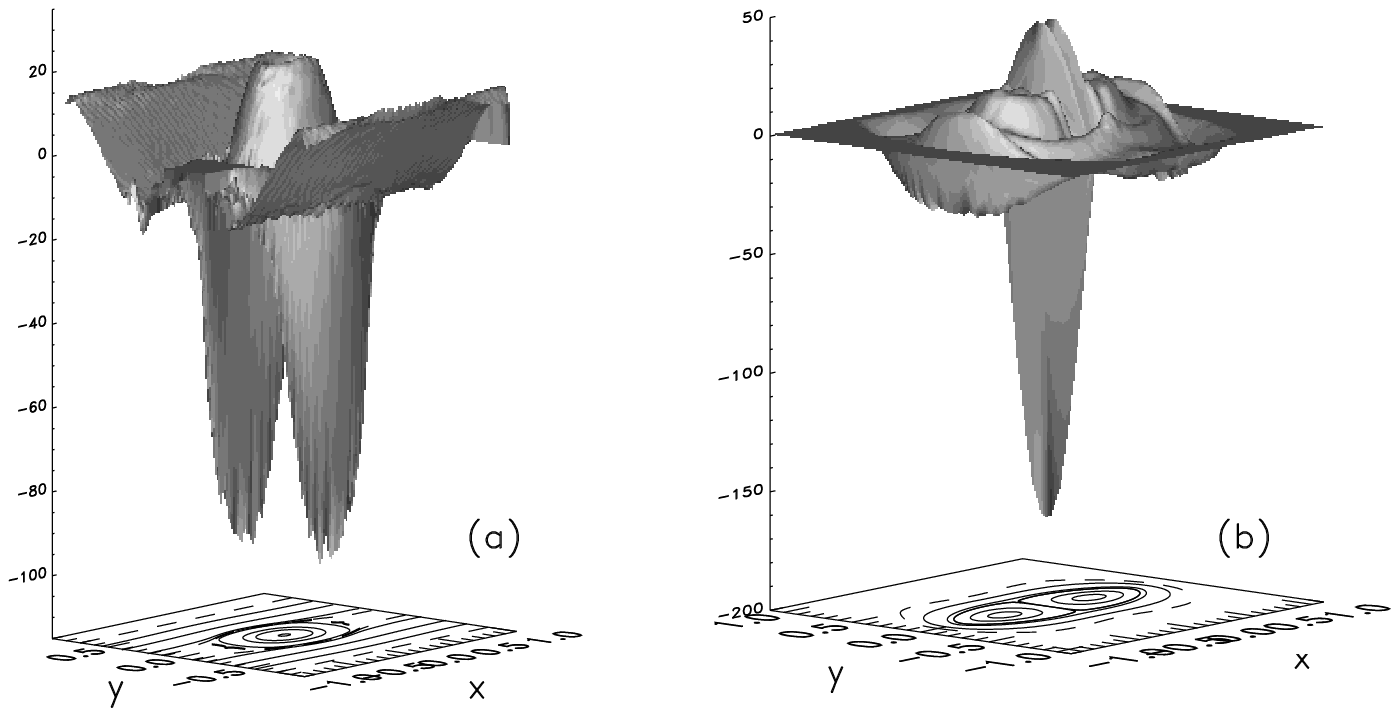,width=3.5in}
$$
\begin{figure}
\caption{Profiles of the $z$-current corresponding to (a) the topology of
Fig.\ 2b, and to (b) Fig.\ 3b (the field line configurations of Figs.~2b and 3b
are reproduced at the bottoms of (a) and (b), respectively). The current sheets
are represented by negative currents. Due to the presence of an external zero
line, the total current is zero, and therefore strong and peaked negative
current is compensated by a spatially distributed positive current. Note
that in panel (a), the current corresponding to the external field of the
magnetic walls changes sign, because each wall contains two zero-lines.
}
\end{figure}
It is crucial to note here that the evolution we just described is not
forced by the walls; thus, the field and fluid near the walls (i.e., on the
wall
side of the zero-lines) has the equilibrium property (\ref{4}). To see
this, note
that on the zero-line, the total pressure is continuous. Thus, we could replace
the initial elliptical field configuration (the ``bubble") lying between
the two
zero-lines with a field-free region whose gas pressure exactly balances the
total
pressure on the wall side of the zero-lines. The resulting configuration is
clearly in equilibrium, and makes clear that the walls do not push the bubble,
i.e., that the evolution of the bubble is entirely driven by the fact that
it is
not in equilibrium. Thus, it is as the bubble tries to become axisymmetric, and
pushes back the walls, that the two discontinuities are formed. In
principle, if
the bubble could reach equilibrium with ellipse-shaped field lines as in Fig.
2(a), then it would not even interact with the walls, and the equilibrium
of the
whole configuration would be smooth. The field evolution to TN here described,
i.e., evolution from an initially smooth state to a state containing a
singularity, is therefore an intrinsic property of the initially smooth state,
rather than being forced by external means.

Consider now the type B configurations. We again conducted two series of
experiments. In the first series of numerical experiments, we studied different
kinds of initial states, with different initial field line shapes, and with
different distributions $A(s)$; in all cases, we again required that the
outermost line must be a zero-line, as shown in Fig.\ 3. We found for the
type B configuration that a field discontinuity always appeared, as in Fig.\ 3b
(which is taken from one of our simulations), no matter what the initial
distribution of $A$, or what kind of analytical representation of the initial
field lines we used.

In the second set of experiments, we simulated the evolution of magnetic field
with different number of field lines. The issue is as follows: The magnetic
field gradient at $x=0$ increases during the evolution, so that the current,
$\nabla\times\bf B$, approaches a $\delta$-function (Fig.\ 5). It is
not possible to observe this tangential discontinuity because in the
simulations, the field is described via a finite (albeit a very large) number
of field lines (recall \cite{resolution}). Our hypothesis is that in the
limit of
an infinite number of field lines $N\to\infty$, the current at $x=0$ tends
to infinity; in order to test this  hypothesis, we increased the number $N$
(recall \cite{number}) in a succession of simulations that were otherwise
identical. According to our hypothesis, we expected the current to grow
roughly as
$1/\Delta$, where $\Delta$ is the closest distance between the $X$-point
and the nearest field line; the experiments confirmed this expectation.

\section{Conclusion.}

The fundamental result emerging from our simulations is that the vanishing
magnetic field on the outermost field line imposes strict constraints on the
geometry of equilibrium: The type A topology can be at equilibrium only if
it is axisymmetric; and therefore, if constrained by external walls, it is
at TN.
Similarly, the type B rosette structure develops discontinuities, but only
in the presence of an external zero-line. The presence of zero-lines is 
thus an important aspect of topological nonequilibria.

Finally, we comment briefly on the applicability of these results to
astrophysical situations. Observations of the solar atmosphere \cite{Shibata93}
commonly show topologically unconnected magnetic flux systems which are seen to
interact (viz., emerging flux loops). In such circumstances, in which one
expects to encounter small but finite resistivity, these flux systems are
initially
unlinked, but as they are pushed together (and begin to reconnect), flux
linkage is expected to occur and to lead to a field topology analogous to that
depicted in Fig.\ 2, or to the generic type B configuration, discussed
here. The
magnetic flux surrounding these two islands would be initially weak, and the
current sheet which is formed is therefore expected to be weak. However, during
the course of reconnection, more flux will be pushed outside the two
islands, thus
accelerating the process of reconnection. This process may therefore be
self-accelerating, resulting in final (spontaneous) reconnection; preliminary
numerical simulations of a resistive case of this sort suggest that the
reconnection rate $v_d$ scales as $c_A/S^\alpha$, where $\alpha$ is a
small power, $\alpha\sim O(0.1)$
\cite{jon}. If confirmed, it would imply that the reconnection is fast
enough to satisfy the observed
(solar) constraints on reconnection times.  (Recall that while the Sweet-Parker
reconnection time for typical parameters corresponding to the solar corona is
about three years, the time corresponding to $v_d=c_A/S^{0.1}$ is only 30
minutes,
which is comparable to the energy release time scale for large solar flares,
related to the so-called ``long-enduring" events \cite{golub}). Therefore
the two
topologies depicted in Figs.\ 2 and 3 may be regarded as generic examples of
``fast" reconnection and activity in magnetically active astrophysical systems.

\acknowledgments
We have benefited considerably from comments by, and discussions with,
H.K.\ Moffatt, E.N. Parker, and T. Emonet. This work has been supported by the
NASA Space Physics Theory Program at the University of Chicago (RR, SIV)
and, in part, by NSF grant ATM-9320575, and NASA contracts NAS5-96081 and
NASW-5017
to SAIC (ZM and JAL).


\begin{references}
\bibitem{plasma}
M.\ N.\ Rosenbluth, R.\ Y.\ Dagazian, and P.\ H.\ Rutherford, Phys.\
Fluids, {\bf 16}, 1894 (1973);
M. N. Rosenbluth and M. N. Bussac, Nucl. Fusion {\bf 19}, 489 (1979).
\bibitem{flares}
E.\ N.\ Parker, {\it Spontaneous Current Sheets in Magnetic Fields}
(Oxford, New York, 1994).
\bibitem{golub}
L. Golub and J.M. Pasachoff, {\it The Solar Corona} (Oxford U.\ Press, Oxford,
1997).
\bibitem{priest}
E. R. Priest, {\it Solar Magnetohydrodynamics} (D. Reidel, Dordrecht,
Holland, 1981).
\bibitem{dynamo}
S. I. Vainshtein, R. Z. Sagdeev, and R. Rosner, Phys. Rev. E, 56, 1605
(1997); S. I. Vainshtein, Phys. Rev. Letters, 80, 4879 (1998).
\bibitem{galeev}
A.\ A.\ Galeev, Space Sci. Rev., {\bf 23}, 411 (1979)
\bibitem{parker}
E. N. Parker, Astrophys. J., {\bf 174}, 499 (1972); E. N. Parker, {\it
Cosmical Magnetic Fields}, (Clarendon, Oxford, 1979).
\bibitem{analogy}
H. K. Moffatt, J. Fluid Mech., {\bf 159}, 359 (1985)
\bibitem{more}
H. K. Moffatt, J. Fluid Mech., {\bf 166}, 359 (1986); H. K. Moffatt, in
{\it Turbulence and Nonlinear Dynamics in MHD Flows}, eds. M.
Meneguzzi, A. Pouquet and P. L. Sulem (Elsevier Science Publishers B.
V., North-Holland, 1989).
\bibitem{Bajer}
K. Bajer and H. K. Moffatt, J. Fluid Mech. {\bf 212}, 337 (1990);
D. Linardatos, J. Fluid Mech. {\bf 246}, 569 (1993); A. Y. K. Chui and
H. K. Moffatt, Proc. R. Soc. Lond. A, {\bf 451}, 609 (1995); A. Y. K.
Chui and H. K. Moffatt, J. Plasma Physics {\bf 56}, 677 (1996).
\bibitem{kadom}
B. B. Kadomtsev, Plasma Phys., {\bf 1}, 710 (1975);
S. I. Vainshtein, {\it Magn. Gidrodin.}, {\bf 3}, 269 (1988).
\bibitem{book}
S. I. Vainshtein, Sov. Phys. JETP, {\bf 59}, 262 (1984);
S. I. Vainshtein and E. N. Parker, Astrophys. J., {\bf 304}, 821
(1986); S. I. Vainshtein, A. M. Bykov, and I. N. Toptygin, {\it
Turbulence, Current Sheets and Shocks in Cosmic Plasma}, (Gordon \&
Breach, Amsterdam, 1993).
\bibitem{Yamada} M. Yamada, H. Ji, T. Carter, R. Kulsrud, Y. Ono,
and F. Perkins,  Phys. Rev Lett., {\bf 78}, 3117 (1997).
\bibitem{zmjl}
Z.\ Miki\'c and J.\ A.\ Linker, Astrophys. J., {\bf 430}, 898 (1994).
\bibitem{porous}
Note that Moffatt and collaborators \cite{Bajer} treat
the relaxation problem in quasi-static porous media, so that their
momentum equation differs from ours.
\bibitem{i}
M. B. Isichenko, {\it Rev. Mod. Phys.}, {\bf 64}, 961 (1992);
M. B. Isichenko and A. V. Gruzinov, {\it Phys. Plasmas}, {\bf 1}, 1802
(1994).
\bibitem{suppression}
S. I. Vainshtein, R. Z. Sagdeev, R. Rosner, and E.-J. Kim, Phys. Rev. E,
53, 4729 (1996).
\bibitem{resolution} In our approach, it is necessary for each Eulerian
cell to contain at least several field lines. Therefore, as we increase the
spatial resolution of the Eulerian grid, we need to correspondingly
increase the number of field lines used for the Lagrangian calculation.
\bibitem{number}
In order to resolve the same field with increased
$N$ adequately, we have to increase the number of Eulerian cells; see
\cite{resolution}.
\bibitem{Shibata93}
K. Shibata, N. Nitta, R. Matsumoto, T. Tajima, T. Yokoyama, T. Hirayama, and H.
Hudson, in {\it X-ray Solar Physics from Yohkoh}, eds. Y. Uchida, T.
Watanabe, K.
Shibata, and H. Hudson (Tokyo: Universal Academic Press Inc.), pp. 29-32
(1993).
\bibitem{jon} S. I. Vainshtein, Z. Miki\'c, and J. A. Linker, in
preparation. The relatively simple geometry of the configuration makes it
possible to study the scaling of the reconnection rate with two decades of
scaling in $S$. Presumably, the reconnection rate substantially exceeds the
Sweet-Parker rate because the current sheet is relatively short.
Low rates of reconnection observed in coalescing magnetic
islands, see, e.g., C. Marliani and H.R. Strauss, Physics of Plasmas, {\bf
6}, 495 (1999), may be attributed to the fact that the configuration is
not at topological nonequilibrium, because there is no outermost zero-line
in the problem, so that the current sheet is instead formed due to
instability.


\end{references}
\end{document}